\documentclass[pra,amssymb,amsmath]{revtex4}

\usepackage{graphicx}
\usepackage{amssymb,amsmath,amsbsy} 

\usepackage[polish,english]{babel}
\usepackage[T1]{fontenc}

\def\B#1{\!\left(#1\right)}
\def\BB#1{\!\left[#1\right]}
\def\BBB#1{\!\left|#1\right|}

\def\be{\begin{equation}}
\def\ee{\end{equation}}

\def\bee{\begin{equation*}}
\def\eee{\end{equation*}}

\begin{document}
\title{Fidelity approach to quantum phase transitions in  quantum Ising model}

\author{Bogdan Damski}

\affiliation{Jagiellonian University, Institute of Physics, 
{\L}ojasiewicza 11, 30-348 Krak\'ow, Poland}

\begin{abstract}
Fidelity approach to quantum phase transitions uses the overlap between ground
states of the system to gain some information about its quantum phases. 
Such an overlap is called fidelity. 
We illustrate how this approach works in the one dimensional quantum Ising model 
in the transverse field. Several closed-form analytical expressions for
fidelity are discussed. An example of what insights fidelity provides into the dynamics of
quantum phase transitions  is carefully described.
The role of fidelity in central spin systems is pointed out.
\\~\\{\tt 
Proceedings of the 50th Karpacz Winter School of Theoretical Physics, Karpacz, Poland, 2-9 March 2014} 
\end{abstract}

\maketitle

\section{Introduction}
\label{sec_In}
Quantum phase transitions happen when small variations of an external parameter
fundamentally  change the ground state properties of the system \cite{Sachdev,SachdevToday}. 
They typically appear when there are competing interactions 
trying to order the sample in different ways and the balance
between them is controlled by an external field.

The fidelity approach to quantum phase transitions uses the overlap 
between  ground states to gain some information about the quantum
phases \cite{Zanardi,GuReview}.
For example, we can define fidelity as 
\be
F\B{g,\delta}=|\langle g-\delta|g+\delta\rangle|,
\label{F1}
\ee
where we assumed that the Hamiltonian $\hat H$ of the system depends on some parameter $g$ 
(e.g. an external field whose variation induces a quantum phase transition), $|g\rangle$ stands 
for the non-degenerate ground state of 
the Hamiltonian $\hat H(g)$, and $\delta$ is a parameter shift.
Besides depending on $g$ and $\delta$, fidelity also depends on $N$, i.e., the
size of the system (number of spins, atoms, etc.).

The fidelity approach to quantum phase transitions is based on the expectation
that fidelity should exhibit a marked drop near the critical point, where 
the ground states of the Hamiltonian are most sensitive to the external field. 
Fidelity has been studied in the following limits: 
\begin{itemize}
\item $\delta\to0$ at the fixed system size $N$: one employs here the Taylor expansion to get 
\be
F\B{g,\delta}=1-2\chi(g)\delta^2 + {\cal O}\B{\delta^4},
\label{Fsus}
\ee
where the linear in $\delta$ term disappears due to the normalization
condition $\langle g|g\rangle=1$. Alternatively, one may note that 
such a term could make $F(g,\delta)>1$ or that it would break the
$\delta\to-\delta$ symmetry of $F(g,\delta)$.
The central object of this expansion is the fidelity susceptibility $\chi(g)$.
Using the scaling theory of quantum phase transitions  \cite{Sachdev,ContinentinoBook}, 
it has been shown that at the critical point $g_c$
in a $d$ dimensional system \cite{ABQ2010,Polkovnikov} 
\be
\chi(g_c)\sim N^{2/d\nu},
\label{sc_near}
\ee
while far away from it 
\be
\chi(g) \sim \frac{N}{\BBB{g-g_c}^{2-d\nu}},
\label{sc_away}
\ee
where $\nu$ is the universal critical exponent describing the divergence of the
correlation length $\xi(g)$ near the critical point, $\xi(g)\sim\BBB{g-g_c}^{-\nu}$.
Fidelity susceptibility can be viewed in two ways. First, it can be treated as
a function whose knowledge leads to the determination of fidelity (\ref{F1}) through Eq.
(\ref{Fsus}). One should remember, however, that it can be used in such a way
only when the ${\cal O}\B{\delta^4}$ term is negligible, which sets conditions 
on the system size $N$,  the distance from the critical point $|g-g_c|$, and the field shift $\delta$ . 
For example,
the expansion  (\ref{Fsus})
always breaks for large enough systems. Second, one can formally define 
fidelity  susceptibility through the relation 
\be
\chi(g) = -\left\langle g\BBB{\frac{d^2}{dg^2}}g\right\rangle,
\label{Dsus}
\ee
and  use it to study the quantum phase transitions 
for any system size $N$ and any    parameter $g$ (see e.g.
Refs.  \cite{GeometricXY,GeometricScaling,Kolodrubetz}, where fidelity susceptibility
gives the diagonal elements of the geometric tensor providing some
information about the  quantum critical points).
Finally, we mention that it has been recently proposed that fidelity
susceptibility can be extracted out from the spectral function \cite{GuExp}, which 
can be measured in standard condensed matter setups.

\item $N\to\infty$ at the fixed field shift $\delta$: this is the limit, where 
the Anderson orthogonality catastrophe \cite{Anderson1967}, i.e., the disappearance of the
overlap between ground states of the thermodynamically large system, 
	happens. 
	We  have argued 	that at the critical point \cite{BDfid1} 
	\be
	\frac{\ln F(g_c,\delta)}{N}\sim-|\delta|^{d\nu},
	\label{lnF}
	\ee
	while far away from it 
	\be
	\frac{\ln F(g,\delta)}{N}\sim-\frac{\delta^2}{\BBB{g-g_c}^{2-d\nu}}
	\label{Faway}
	\ee
	as long as $d\nu<2$. Eq. (\ref{Faway}) becomes equivalent to Eq.
	(\ref{sc_away}) when $N\delta^2/\BBB{g-g_c}^{2-d\nu}\ll 1$.
	We mention that $-\ln F/N$ is called fidelity per
	lattice site \cite{Zhou2008,Zhou2008_Ising}.
\end{itemize}

The outline of these lecture notes is the following. Sec. \ref{sec_Qu}
presents basic concepts associated with the quantum Ising model. 
The main stress in this section is placed on
the discussion of the  parity of the ground state of the Ising chain. 
In Sec. \ref{sec_Sus}, we discuss the exact closed-form expression for 
fidelity susceptibility of the Ising model. 
Sec. \ref{sec_The} is devoted to the studies of fidelity in thermodynamically large systems.
We show there how the Anderson orthogonality catastrophe happens in the Ising
chain. Sec. \ref{sec_Dyn} focuses on the dynamics of quantum phase
transitions. We explain there  how fidelity can be used to
obtain the probability of finding the system in the ground state after the
non-equilibrium crossing of the quantum critical point. Finally, 
we briefly discuss in Sec. \ref{sec_Cen} two problems associated with the central spin coupled to the
Ising chain. Their  solution also involves  fidelity.

\section{Quantum Ising model}
\label{sec_Qu}
\begin{figure}[t]
\begin{center}
\includegraphics[width=4.5in]{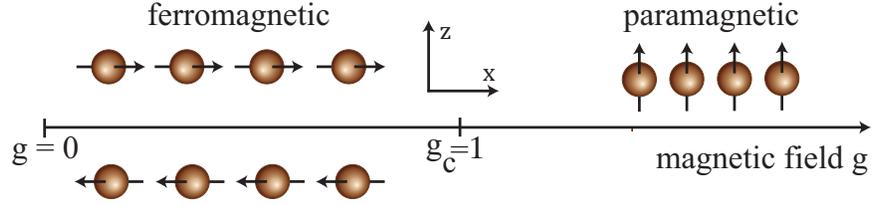}
\end{center}
\caption{Schematic illustration of the quantum phase transition in the
quantum Ising model. Adapted  from Ref.  \cite{BDfid3}.}
\label{f_ising}
\end{figure}

The Hamiltonian of the one dimensional quantum Ising model in the transverse
field reads
\be
\hat H(g) = -\sum_{i=1}^N \B{\sigma^x_i\sigma^x_{i+1} + g\sigma^z_i},
\label{HIsing}
\ee
where $g$ stands for an external magnetic field acting on the system and $N$
is the number of spins. The periodic boundary conditions are assumed, i.e., 
$\sigma^x_{N+1}=\sigma^x_1$.
While the spin-spin interactions in this model try to align the spins
in the $\pm x$ direction, the magnetic field tries to put the spins along
its direction, i.e., $+z$ for $g>0$ and $-z$ for $g<0$. 
The competition between different spin orientations results in the quantum
phase transition between the ferromagnetic and  paramagnetic phases (Fig.
\ref{f_ising}). 
The system is in the ferromagnetic phase for $|g|<1$, it is in the
paramagnetic phase for $|g|>1$, and it is at the quantum critical point 
when  $|g|=1$, i.e., $g_c=\pm1$. The experimental possibilities for the studies of this model 
and other  Ising-like models   appear in  cobalt niobate \cite{ColdeaSci2010}, as
well as  in cold ion \cite{MonroeAll,LanyonSci2011}, cold atom \cite{GreinerNat2011}, and NMR \cite{LiPRL2014}
simulators of spin systems.

This model is exactly solvable \cite{Lieb1961,Pfeuty,BarouchPRA1971}. Its role in
the studies of strongly correlated quantum systems parallels  the role of the harmonic
oscillator in the single-particle quantum theory: 
Numerous theoretical concepts in many-body quantum physics are being tested
in the quantum Ising chain in the transverse field.

To correctly compute fidelity of the quantum Ising model, one has to  
properly identify the parity symmetry of its ground state \cite{ABQ2010,BDfid4}. 
We will follow below the discussion from Ref.  \cite{BDfid4}, where the 
comprehensive study of this problem in a {\it finite} Ising chain is
presented (see also Ref.  \cite{XY} for the discussion of the finite-size effects
in the XY model). 

First, we note that the Hamiltonian (\ref{HIsing}) commutes with the parity operator
\bee
[\hat H,\hat P]=0, \ \ \hat P =  \prod_{i=1}^N \sigma^z_i,
\eee
whose eigenvalues are either $+1$ or $-1$. Thus, the Hamiltonian (\ref{HIsing}) can be 
independently diagonalized in the positive and negative parity subspaces.

Second, one performs the Jordan-Wigner transformation mapping the spin-1/2
operators to fermionic operators
\bee
\sigma_i^z=1-2\hat c_i^\dag \hat c_i, 
\ \sigma_i^x=\B{\hat c_i + \hat c_i^\dag}\prod_{j<i}\B{1-2\hat c_j^\dag\hat c_j}, 
\  \{\hat c_i,\hat c_j^\dag\} = \delta_{ij}, 
\ \{\hat c_i,\hat c_j\} = 0.
\eee
Substituting this into Eq. (\ref{HIsing}),  one obtains 
\be
\begin{aligned}
&\hat H(g) = -\sum_{i=1}^{N} \BB{\hat f_{i,i+1} + g\B{\hat c_i \hat c_i^\dag - \hat c_i^\dag \hat c_i}},\\
&\hat f_{i,j} = 
\hat c_i^\dag\hat c_j - \hat c_i \hat c_j^\dag - \hat c_i\hat c_j +  \hat c_{i}^\dag \hat c_j^\dag,
\end{aligned}
\label{Hc}
\ee
where 
\be
\hat c_{N+1}=-\hat c_1
\label{canty}
\ee
should be taken in the positive parity subspace and 
\be
\hat c_{N+1}=\hat c_1
\label{cper}
\ee
in the negative parity subspace. 
Since the Hamiltonian (\ref{Hc}) is quadratic in the fermionic operators, it can
be diagonalized in the standard way. One first  substitutes   
\be
\hat c_j= \frac{\exp(-i\pi/4)}{\sqrt{N}}\sum_{K=\pm k} \hat c_K\exp(iKj)
\label{qwerty}
\ee
into the Hamiltonian (\ref{Hc}),
where\footnote{Whenever $k=0,\pi$, one should keep only the $K=+k$ term
in the sum (\ref{qwerty}). 
}
\be
\begin{aligned}
&k=\frac{\pi}{N},\frac{3\pi}{N},\frac{5\pi}{N},\cdots,\pi-\frac{\pi}{N} \ \ {\rm for \ even} \ N,\\
&k=\frac{\pi}{N},\frac{3\pi}{N},\frac{5\pi}{N},\cdots,\pi \ \ {\rm for \ odd} \ N,
\end{aligned}
\label{kplus}
\ee
in the positive parity subspace and  
\be
\begin{aligned}
&k=0,\frac{2\pi}{N},\frac{4\pi}{N},\cdots,\pi \ \ {\rm for \ even} \ N,\\
&k=0,\frac{2\pi}{N},\frac{4\pi}{N},\cdots,\pi-\frac{\pi}{N} \ \ {\rm for \ odd} \ N,
\end{aligned}
\label{kminus}
\ee
in the negative parity subspace. 
The quantization of momenta  (\ref{kplus}) and (\ref{kminus}) follows from the boundary conditions  
(\ref{canty}) and (\ref{cper}), respectively.
Then one performs the standard  Bogolubov transformation to diagonalize the resulting Hamiltonian.

Third, one needs to determine whether the ground state of the Ising Hamiltonian
(\ref{HIsing}) lies in the positive or negative parity subspace. This can be 
done by computing the lowest eigenenergy in each of the subspaces, say
$\varepsilon^+$ and $\varepsilon^-$ in the positive and negative parity subspaces,
respectively. We find
that $\varepsilon^--\varepsilon^+$ equals \cite{BDfid4} 
\bee  
g^N \int_0^1 dt\, \frac{4 N}{\pi} \frac{t^{N-3/2} \sqrt{(1-t)(1-g^2 t)} }{1-(gt)^{2N} }
\eee
in the ferromagnetic phase,
\bee
{\rm sign}\B{g^N} \B{2|g|-2} +   g^{-N} \int_0^1 dt\,\frac{4 N}{\pi} \frac{t^{N-3/2} \sqrt{(1-t)(g^2-t)} }{1- t^{2N}/g^{2N}}
\eee
in the paramagnetic phase, and  
\bee
2\tan\B{\frac{\pi}{4N}} {\rm sign}\B{g_c^N} 
\eee
at the critical points.

A quick look at these expressions shows that 
the parity of the ground state of the Hamiltonian (\ref{HIsing}) is
\begin{itemize}
\item positive for all magnetic fields when the number of spins  $N$ is even, 
\item positive for $g>0$ and negative for $g<0$  in the odd-sized systems.
\end{itemize}
These remarks follow from the observation that 
${\rm sign} \B{\varepsilon^- - \varepsilon^+} = {\rm sign}\B{g^N}$.
The same result  was obtained in a different way in Ref.  \cite{ABQ2010}
for $g>0$.

Besides providing the parity of the ground state, these expressions quantify
the gap between the eigenenergies of the positive and negative 
parity ground states of the Hamiltonian (\ref{HIsing}):
\begin{itemize}
\item {\it Ferromagnetic phase} ($|g|<1$). The gap is exponentially small 
in the  thermodynamically large systems 
\bee
\left|\varepsilon^--\varepsilon^+\right| = {\cal O}\B{\exp\B{-N/\xi(g)}/\sqrt{N}},
\eee
where 
\be
\xi(g)=1/\BBB{\ln\BBB{g}}
\label{xii}
\ee
is the  correlation length of the {\it infinite} Ising chain \cite{BarouchMcCoy1971}
and we define the thermodynamic limit by the relation 
\be
N\gg\xi(g).
\label{NggXi}
\ee
In the opposite limit, 
\bee
N\ll\xi(g),
\eee
we find  that the gap is polynomial in the system size
\bee
\left|\varepsilon^- - \varepsilon^+\right| = {\cal O}\B{1/N}.
\eee
\item {\it Paramagnetic phase} ($|g|>1$). The gap is always macroscopic and grows
approximately linearly with the distance from the critical point.
\item {\it Critical points} $|g|=1$. The gap is inversely proportional to the
size of the system in large systems.
\end{itemize}
At the risk of stating the obvious, we mention that the {\it qualitative} content of
these remarks is discussed in textbooks on quantum phase transitions (see e.g.
Ref.  \cite{Sachdev}).

\section{Fidelity susceptibility of  quantum Ising model}
\label{sec_Sus}
Our discussion in this section will be based on Refs.  \cite{BDfid3} and
 \cite{BDfid4}. The former one  provides the exact closed-form expression 
for fidelity susceptibility of even-sized Ising chains and studies  its basic
properties. The latter one
generalizes this result to the odd-sized  chains and discusses the
pitfalls of incorrect identification of the parity of the ground state  in 
fidelity susceptibility computations.

The ground state wave-function of the Ising chain can be written as 
\be
\begin{aligned}
|g\rangle &= \prod_{k} 
 \B{\cos\B{\frac{\theta_k}{2}} -
\sin\B{\frac{\theta_k}{2}}\hat c_k^\dag\hat c_{-k}^\dag}|{\rm vac}\rangle, \\
\sin\B{\theta_k} &= \frac{\sin(k)}{\sqrt{g^2-2g\cos(k)+1}}, \ \ 
\cos\B{\theta_k}= \frac{g-\cos(k)}{\sqrt{g^2-2g\cos(k)+1}},
\end{aligned}
\label{GSref}
\ee
where the proper momenta $k$ have to be used: (\ref{kplus}) if the ground
state of the Hamiltonian (\ref{HIsing}) lies in  the positive
parity subspace and (\ref{kminus}) if it belongs to the negative parity
subspace (see Ref.  \cite{BDfid4} for the discussion of the $k=0,\pi$ modes that do not contribute to the 
following calculations).
  The $|{\rm vac}\rangle$ state is annihilated by all $\hat c_k$ operators.

Using this wave-function one easily finds with Eq. (\ref{Dsus}) that 
\be
\chi(g)=\frac{1}{4}\sum_{k}\B{\frac{d\theta_k}{dg}}^2=
\frac{1}{4}\sum_{k} \frac{\sin^2\B{k}}{\B{g^2-2g\cos\B{k}+1}^2}.
\label{ifidsus}
\ee
This summation has been analytically performed in Refs.  \cite{BDfid4} and  \cite{BDfid3}.

\begin{figure}[t]
\begin{center}
\includegraphics[width=4.5in]{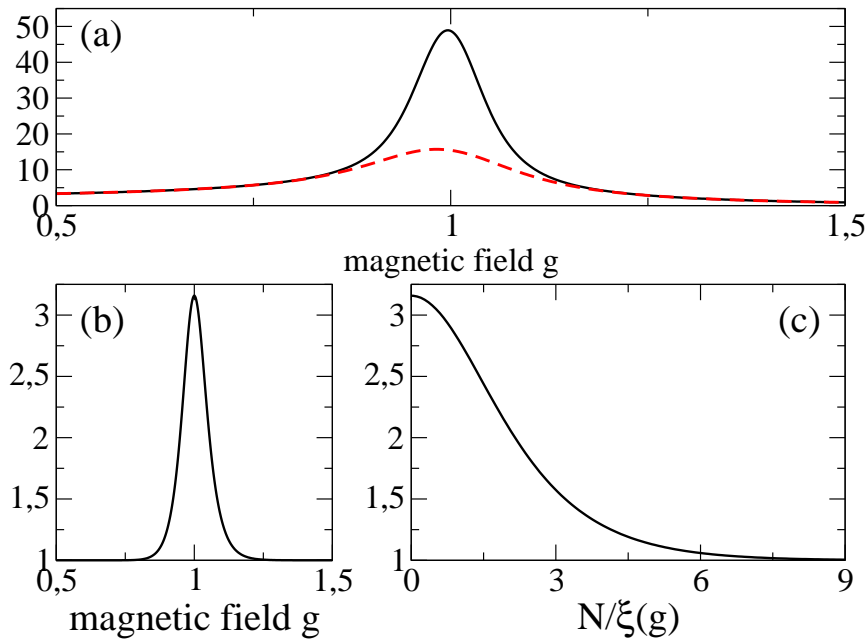}
\end{center}
\caption{Panel (a): black solid line shows $\chi^+$ (\ref{chipositive}), while the red dashed
line shows $\chi^-$ (\ref{chinegative}). Even $N=40$ is used. Panel (b): 
The $\chi^+/\chi^-$ ratio for data from the (a) panel. Panel (c): the same as
panel (b), but as a function of the ratio between the system size and the
correlation length $N/\xi(g)=N|\ln\BBB{g}|$.}
\label{fig_pm0}
\end{figure}

\begin{figure}[t]
\begin{center}
\includegraphics[width=4.5in]{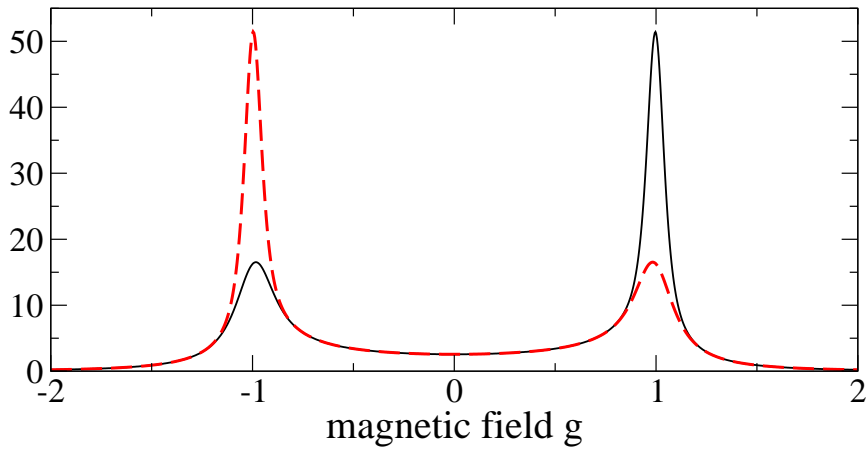}
\end{center}
\caption{Black solid line shows $\chi^+$ (\ref{chipositive}), while the red dashed
line shows $\chi^-$ (\ref{chinegative}). The plot is done for odd $N=41$.}
\label{fig_pm}
\end{figure}

Before providing the definite closed-form expression for fidelity susceptibility,
we emphasize the need for the correct identification of the parity of the ground
state \cite{BDfid4}. 
If we use the momenta (\ref{kplus}) to compute the series (\ref{ifidsus}), we will obtain 
\be
\chi^{+}(g)= 
\frac{N^2}{16g^2}\frac{g^N}{\B{g^N+1}^2}+\frac{N}{16g^2}
\frac{g^N-g^2}{\B{g^N+1}\B{g^2-1}}.
\label{chipositive}
\ee
If we, however, sum the series (\ref{ifidsus}) over momenta
(\ref{kminus}), we will get
\be
\chi^{-}(g)= 
-\frac{N^2}{16g^2}\frac{g^N}{\B{g^N-1}^2}+\frac{N}{16g^2}
\frac{g^N+g^2}{\B{g^N-1}\B{g^2-1}}.
\label{chinegative}
\ee
These two expressions are plotted in Fig. \ref{fig_pm0}a. 
They differ negligibly away from the critical point, i.e.,
for magnetic fields $g$ such that $N\gg\xi(g)$. 
On the other hand, near the critical point, i.e., for magnetic fields $g$ such that 
the system size $N$ is smaller than a few correlation lengths $\xi(g)$, 
they substantially differ (Fig. \ref{fig_pm0}c).  
This difference was   numerically
studied in Ref.  \cite{ABQ2010}, where it caused problems in the Quantum Monte Carlo
studies of fidelity susceptibility at non-zero temperature. 
Our expressions analytically  quantify this difference in the zero temperature limit.
For example,  
\begin{itemize}
\item right at the critical point $g_c=1$ the ratio of the two expressions equals
      \bee
      \left.\frac{\chi^+}{\chi^-}\right|_{g_c=1} = 3 \frac{N}{N-2},
      \eee
      and so it  approaches $3$ as $N$ increases (Figs. \ref{fig_pm0}b
      and \ref{fig_pm0}c),
\item in odd-sized systems neither of the two expressions, (\ref{chipositive})
and (\ref{chinegative}), is symmetric with respect to the $g\leftrightarrow-g$ 
transformation (Fig. \ref{fig_pm}). 
\end{itemize}

Using the proper identification of the parity of the ground state 
discussed in  Sec. \ref{sec_Qu}, one quickly finds that 
(i) $\chi(g)=\chi^+(g)$ for all system sizes and $g>0$; (ii)
 $\chi(g)=\chi^+(g)$ for even-sized systems and $g<0$; and (iii)
 $\chi(g)=\chi^-(g)$ for odd-sized systems and $g<0$. Correlating the  
 observations (i) -- (iii) with equations (\ref{chipositive}) and
 (\ref{chinegative}), we get 
\be
\chi(g)= 
\frac{N^2}{16g^2}\frac{|g|^N}{\B{|g|^N+1}^2}+\frac{N}{16g^2}
\frac{|g|^N-g^2}{\B{|g|^N+1}\B{g^2-1}}.
\label{chifinal}
\ee
This exact closed-form expression works for any system size $N\ge2$ and any magnetic field $g$.
As expected, it is symmetric with respect to the $g\leftrightarrow-g$
transformation. Its properties  can be summarized as follows \cite{BDfid3}.

First, it undergoes  a simple transformation law under the Kramers-Wannier
duality mapping \cite{Kramers}
\be
g \leftrightarrow \frac{1}{g}.
\label{KM}
\ee
Indeed, one easily verifies that \cite{BDfid3}
\be
g^2\chi(g)=\B{\frac{1}{g}}^2\chi\B{\frac{1}{g}}.
\label{symmetry}
\ee

Second, one  finds that away from the critical point
in the paramagnetic phase 
\be
\chi(g) = \frac{N}{16g^2\B{g^2-1}} + R\B{|g|}, 
\label{away1}
\ee
where the remainder $R$, which can be easily computed from Eq. (\ref{chifinal}),
can be shown \cite{BDfid3} to be negligible when inequality (\ref{NggXi}) holds.
It should be stressed that the main part of this result can be obtained by  
replacing the sum by the integral (see e.g. Refs. 
 \cite{GuReview},
 \cite{Polkovnikov}, and 
 \cite{GeometricXY})
\be
\sum_k \to \frac{N}{2\pi}\int_0^\pi dk
\label{sumint}
\ee
in Eq. (\ref{ifidsus}). While such a replacement works well away from the critical point, it leads 
to  a completely wrong result near the critical point. The formal approach to 
the replacement of the sum by the integral is provided by the Euler-Maclaurin
summation formula  \cite{Watson}. For example, adopting this formula to the even $N$
positive parity case, we get
\be
\begin{aligned}
\sum_{k=\pi/N}^{\pi-\pi/N}f(k)&=\frac{N}{2\pi}\int_{\pi/N}^{\pi-\pi/N}dkf(k) \\
                              &+\frac{1}{2}\BB{f\B{\frac{\pi}{N}}+f\B{\pi-\frac{\pi}{N}}}\\
			      &+\sum_{m=1}^{s-1}\frac{(-1)^m B_m}{(2m)!}\B{\frac{2\pi}{N}}^{2m-1}
			      \BB{f^{(2m-1)}\B{\frac{\pi}{N}}-f^{(2m-1)}\B{\pi-\frac{\pi}{N}}}\\
			      &-\B{\frac{2\pi}{N}}^{2s}\frac{1}{(2s)!}
			      \int_0^1dt\,\phi_{2s}(t)\sum_{k=\pi/N}^{\pi-3\pi/N}
			      f^{(2s)}\B{k+t\frac{2\pi}{N}},
\end{aligned}
\label{EM}
\ee
where 
\bee
f(k)=\frac{1}{4}\frac{\sin^2\B{k}}{\B{g^2-2g\cos\B{k}+1}^2}, \ \
f^{(m)}(k)=\frac{d^m}{dk^m}f(k), 
\eee
an integer $s\ge1$
is a parameter that can be chosen at will, $B_s$ are Bernoulli numbers, 
and $\phi_s$ are  the Bernoulli polynomials (we use the convention for $B_s$ and
$\phi_s$ from Ref.  \cite{Watson}). 
Therefore, by performing (\ref{sumint}) one skips the second, third and the
fourth line of the right hand side of Eq. (\ref{EM}), and 
replaces the integration ranges
$\pi/N$ and $\pi-\pi/N$ in the first line by  $0$ and $\pi$, respectively.
This procedure is bound to fail near the critical point.
A quick inspection of expression
(\ref{EM}) shows that the exact computation of  the 
sum (\ref{ifidsus}) is easier than the formal study of the Euler-Maclaurin expression (\ref{EM})!

Before moving on, we mention that the application of the duality mapping (\ref{KM})
to Eq. (\ref{away1}) provides us with the following expression for fidelity 
susceptibility in the ferromagnetic phase 
\be
\chi(g) =\frac{N}{16\B{1-g^2}} + \frac{R\B{1/|g|}}{g^4},
\label{away2}
\ee
where again the remainder is negligible away from the critical point,
i.e., when the condition (\ref{NggXi}) is satisfied.

Third, near the critical point one can expand the exact expression 
in  a series to get \cite{BDfid3}  
\bee
\chi(g) = \frac{N(N-1)}{32g^2}\B{1-\frac{N+1}{N} \frac{(N\ln|g|)^2}{6} + r\B{|g|}}, 
\eee
where this time the reminder $r(|g|)$ is negligible for $N\ll\xi(g)$. In
particular, at the critical points one finds 
\be
\chi(g_c) = \frac{N(N-1)}{32}.
\label{at}
\ee

Fourth, noting that $d=\nu=1$ and $g_c=\pm1$ in our quantum Ising model \cite{BarouchPRA1971}, 
one easily verifies the scaling relations (\ref{sc_near}) and (\ref{sc_away}) by comparing them to 
the equations  (\ref{away1}), (\ref{away2}), and (\ref{at}). 

Fifth, one can show from the exact solution (\ref{chifinal}), that the maximum of
fidelity susceptibility is located in the ferromagnetic phase  at the distance 
\bee
\frac{6}{N^2}-\frac{6}{N^3} + {\cal O}\B{N^{-4}}
\eee
from the critical point \cite{BDfid3}.
The fact that the maximum is shifted away from the critical point due to the finite-size
effects is not surprising: The shift proportional to $1/N^2$ is 
expected for free fermionic systems \cite{ZanardiMax}. The shift of the maximum
into the ferromagnetic phase is the consequence of the duality symmetry
(\ref{symmetry}), which the reader can easily verify. 

\section{Fidelity in  thermodynamically large Ising chain}
\label{sec_The}
Using wave-function (\ref{GSref}), one easily finds that 
\be
\begin{aligned}
&F(g,\delta)=\prod_k \cos\B{\frac{\theta_+(k)-\theta_-(k)}{2}},\\
&\tan\theta_\pm(k)=\frac{\sin(k)}{g\pm\delta-\cos(k)}.
\end{aligned}
\label{Ftherm}
\ee
We are now interested in studying this expression in the thermodynamic limit near
the critical point $g_c=1$. In fact, 
from now on we will focus on $g\ge0$ and assume even $N$  for convenience.
This section is based on our Refs.  \cite{BDfid1} and  \cite{BDfid2}.

\begin{figure}[t]
\begin{center}
\includegraphics[width=4.5in]{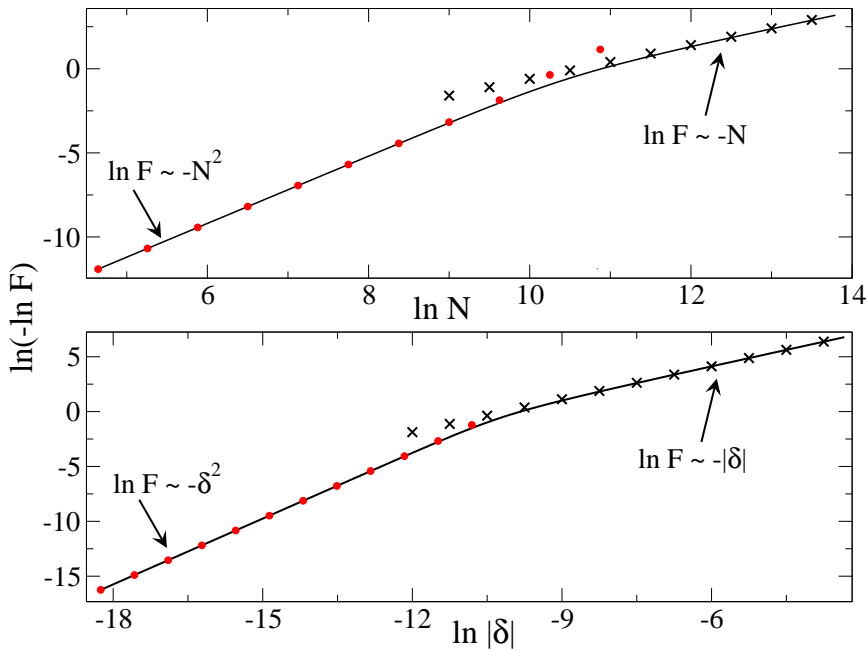}
\end{center}
\caption{Illustration of the transition to the thermodynamic limit. In both
panels the black solid  line shows $F(g_c=1,\delta)$. Upper panel:
the field shift $\delta=10^{-4}$. Lower panel: the
system size $N$ is set to $10^5$. Red dots come from $F=1-N(N-1)\delta^2/16$, i.e., 
the approximation of fidelity (\ref{Fsus}) through fidelity susceptibility (\ref{at}). The black
crosses come from the leading-order thermodynamic approximation  (\ref{lnFcfinal}), 
which for $g=1$, i.e., $c=0$, reads $\ln F=-N|\delta|/4$.}
\label{fig_thermo}
\end{figure}

The transition to the thermodynamic limit can be induced
by either the change of the system size $N$ or the change of the field shift
$\delta$ (Fig. \ref{fig_thermo}). It happens in our problem when \cite{BDfid1}
\bee
N\gg\min[(\xi(g+\delta), \xi(g-\delta)].
\eee
Substituting $\xi(g\pm\delta)$ (\ref{xii}), we find that near
the critical point this condition implies that the system reaches the thermodynamic
limit when  
\be
N|\delta|\gg1,
\label{poi}
\ee
which is numerically studied in Ref.  \cite{BDfid1}. It explains why the
change of either $N$ or $\delta$ can drive the system into the thermodynamic
limit.

It is currently unknown how to find a closed-form expression for the product (\ref{Ftherm}). Thus, we have
to restore to approximations. We assume that $N\to\infty$ at the fixed $\delta$, 
take the logarithm of Eq. (\ref{Ftherm}), and then replace the sum over momenta with the 
integral (\ref{sumint}). 

Introducing  the relative distance from the critical point 
\bee
c=\frac{g-1}{|\delta|},
\eee
we obtain the following approximate result \cite{BDfid1,BDfid2}
\be
\frac{\ln F(g,\delta)}{N} \simeq -|\delta| A(c),
\label{lnFcfinal}
\ee
where 
\be
\begin{aligned}
&A(c)= \left \{
\begin{array}{cc} 
\begin{split}
& \frac{1}{4}+ \frac{|c| K(c_1)}{2\pi}+\frac{(|c|-1) {\rm Im} E(c_2)}{4\pi}& {\rm for}~ |c|<1,  \\
& \frac{|c|}{4} - \frac{|c| K(c_1)}{2\pi}-\frac{(|c|-1) {\rm Im} E(c_2)}{4\pi}& {\rm for}~ |c|\ge1,
\end{split}
\end{array}
\right. \\
&c_1 = -4\frac{|c|}{(|c|-1)^2}, \ \ c_2=\frac{(|c|+1)^2}{(|c|-1)^2}.
\end{aligned}
\label{Acc}
\ee
The complete elliptic integrals of the first and second
kind are defined as
\bee
K(x)=\int_0^{\pi/2} \frac{d\phi}{\sqrt{1-x \sin^2\B{\phi}}}, \ \  
E(x)=\int_0^{\pi/2} d\phi \sqrt{1-x \sin^2\B{\phi}},
\eee
respectively. Several remarks are in order now.

\begin{figure}[t]
\begin{center}
\includegraphics[width=4.5in]{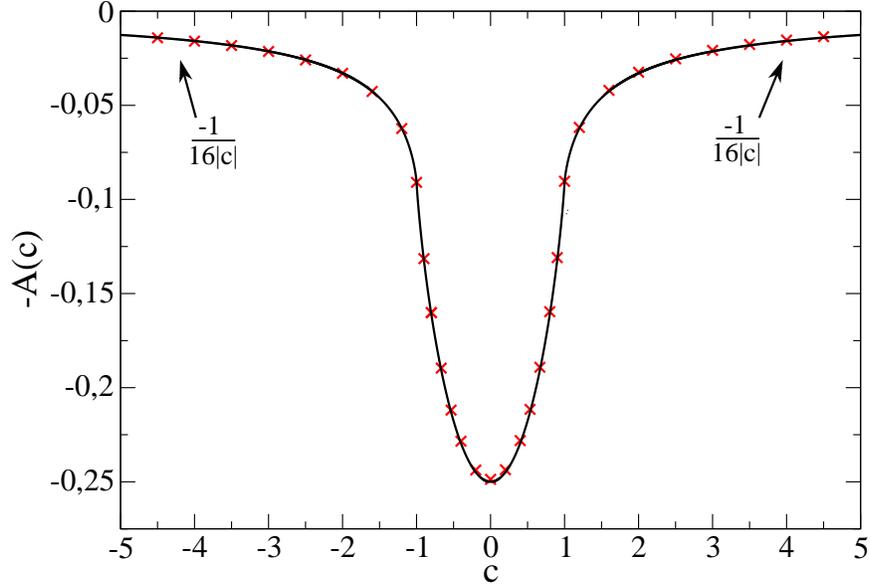}
\end{center}
\caption{The scaling function $A(c)$ (\ref{Acc}). Black line shows $-A(c)$ while the red crosses show 
$\ln F\B{1+c\BBB{\delta},\delta}/N|\delta|$, where $N=10^5$ and
$\delta=\pi/1000$. Note that the thermodynamic limit condition is  well-reached for such a 
system as $N|\delta|=100\pi\gg1$. Adapted   from Ref.
 \cite{BDfid1}.
}
\label{fig_Ac}
\end{figure}

First,  Eq. (\ref{lnFcfinal}) 
provides  the leading order term of the $N\to\infty$ expansion of the exact expression for fidelity per 
lattice site, i.e., $-\ln F(g,\delta)/N$. Its accuracy is illustrated in Fig. \ref{fig_Ac}. 
The next order correction to fidelity per lattice site is discussed in Ref.
 \cite{BDfid2}, where we refer the reader for the details. It comes from the difference between 
the  sum and the integral:
\be
\sum_{k=\pi/N}^{\pi-\pi/N} 
\ln\cos\B{\frac{\theta_+(k)-\theta_-(k)}{2}}
- \frac{N}{2 \pi} \int_0^\pi dk\,\ln\cos\B{\frac{\theta_+(k)-\theta_-(k)}{2}}.
\label{ghjk}
\ee
In particular, the next order approximation to $\ln F(g,\delta)/N$ at the quantum critical point $g_c=1$ is
\be
\frac{\ln F(1,\delta)}{N}\simeq-\frac{|\delta|}{4}+\frac{\ln\B{2}}{2N},
\label{lnFc1}
\ee
where the $\ln\B{2}/2$ prefactor of the subleading term comes from the evaluation of Eq. (\ref{ghjk})
in the $N\to\infty$ limit \cite{BDfid2}.
We will show in Sec. \ref{sec_Dyn} that the  subleading term 
can be accurately extracted out from the probability 
of finding the system in the ground state after the non-equilibrium 
quantum phase transition.

Second, it should be stressed that Eq. (\ref{lnFcfinal}) is derived for
$|\delta|\ll1$ and $|g-1|=|c\delta|\ll1$. This is sufficient for 
describing the sharp drop of fidelity in the vicinity of the critical point.
It should be noticed that fidelity in the
thermodynamic limit can change by several orders of magnitude around the
critical point when the magnetic field is changed by a few $\delta$'s.
Note that this observation holds for $1/N\ll\BBB{\delta}\ll1$, so fidelity is indeed 
a sensitive probe of quantum criticality!

Third, Eq. (\ref{lnFcfinal})  analytically shows 
how the Anderson orthogonality catastrophe happens in the quantum Ising model. 
It proves that the overlap between the Ising model ground states in the quantum critical region 
is exponentially  small in the system size in the thermodynamic limit.

Fourth, the expression for $A(c)$ can be simplified for $|c|\gg1$. Indeed,  one
finds that $A(c)$ approaches $1/16|c|$ in this limit (Fig. \ref{fig_Ac}). This leads to 
the following approximation
\be
F(g,\delta)\simeq\exp\B{-\frac{N\delta^2}{16|g-1|}}.
\label{zxcv}
\ee
If additionally $-N\delta^2/16|g-1|\ll1$, then this expression yields the same result as  
Eq. (\ref{Fsus}). 
Finally, we note that the solutions (\ref{lnFcfinal}) and (\ref{zxcv}) agree with the scaling results 
(\ref{lnF}) and (\ref{Faway}), respectively.

\section{Dynamics of quantum phase transitions}
\label{sec_Dyn}
We will discuss now what insights into the dynamics of the quantum quench 
are provided by fidelity. Although the discussion below is based on Sec. V of Ref.
 \cite{BDfid2},  the numerical results presented here  have not
been published before. We focus here on the dynamics of the quantum Ising
chain, but the discussion can be easily generalized to other ``typical'' quantum critical
systems (see the end of this section).

To start, we consider an instantaneous quench, i.e., a sudden change of the
parameter $g$ of the Hamiltonian (\ref{HIsing}) from $g_1$ to $g_2$. 
If the system was initially prepared in the ground state $|g_1\rangle$,
then the probability of finding it in the ground state after the quench equals 
\bee
\BBB{\langle g_1|g_2\rangle}^2 = F^2(g,\delta), \ \ g=\frac{g_1+g_2}{2}, \  \delta=\frac{g_2-g_1}{2},
\eee
where $F(g,\delta)$ is given by Eq. (\ref{Ftherm}) and the approximations discussed in Secs.
\ref{sec_Sus} and \ref{sec_The} can be used in the appropriate parameter ranges.

\begin{figure}[t]
\begin{center}
\includegraphics[width=4.5in]{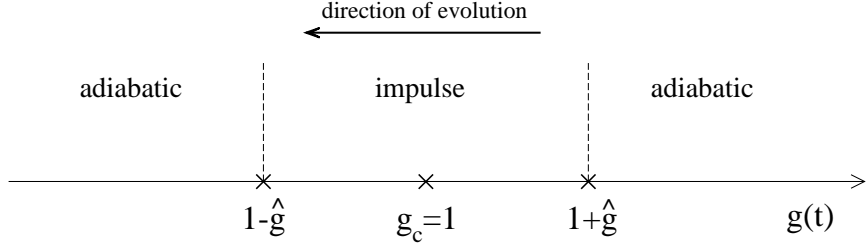}
\end{center}
\caption{Illustration of the adiabatic-impulse approximation in the quantum
Ising chain.}
\label{fig_aia}
\end{figure}

\begin{figure}[t]
\begin{center}
\includegraphics[width=4.5in]{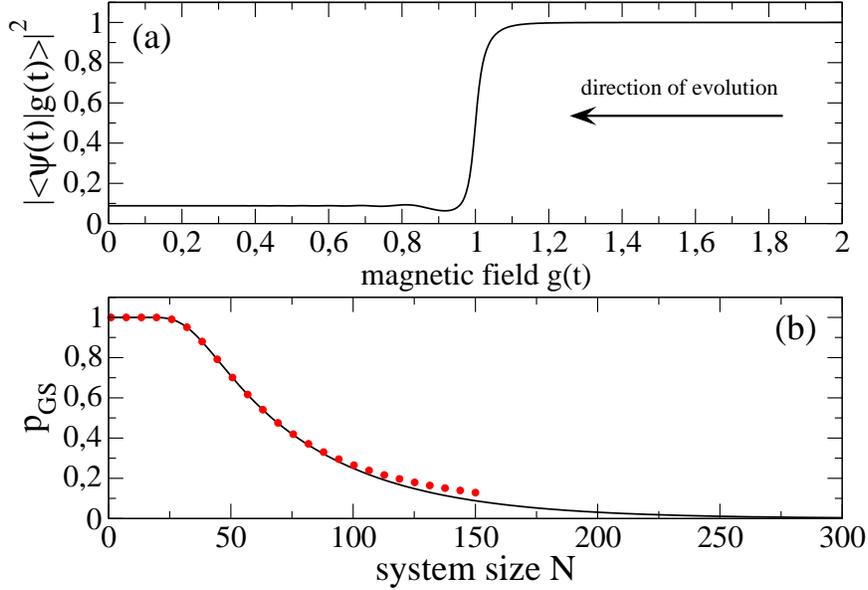}
\end{center}
\caption{Panel (a): the probability of finding the $N=150$ Ising chain 
in the {\it instantaneous} ground state during a quench.
Panel (b): the probability $p_{GS}$ of finding the Ising chain in the ground state by the {\it end}
of time evolution.
The quench protocol in both panels is provided by Eq. (\ref{godt}) with $\tau_Q=50$. 
The evolution starts from the paramagnetic
phase  ground state at the magnetic field $g(t=-5\tau_Q)=5$ and it ends in the ferromagnetic phase 
at the magnetic field $g(t=0)=0$. The solid line on
both panels is a numerical calculation.
The red dashed-dotted line in the lower  panel
shows $1-\exp\B{-2\pi^3\tau_Q/N^2}$. It illustrates  what happens when
the finite-size effects play a role \cite{Marek_decoh}. 
}
\label{fig_prob}
\end{figure}

A much more interesting quench dynamics shows when the magnetic field is continuously changed in time
\be
g(t) = -\frac{t}{\tau_Q},  
\label{godt}
\ee
where $t:-\infty\to0$ and the quench time $\tau_Q$ controls the speed of the driving.
The dynamics of the Ising chain due to such time variation of the magnetic
field is exactly solved in Ref.  \cite{JacekPRL} 
(see Refs.  \cite{JacekAdv} and   \cite{PolkovnikovRMP2011} 
for the recent reviews of the dynamics of quantum phase transitions).
We will now link fidelity to the probability of finding the system in the ground state
after such a quench. To do so,  we invoke the quantum version of the adiabatic-impulse
approximation
discussed  in Ref.  \cite{BDPRL2005} 
(see also Ref.  \cite{DornerPRL2005}).
This approximation splits the evolution across the critical point 
into three parts depicted in Fig. \ref{fig_aia}:
\begin{itemize}
\item The first adiabatic regime takes place when  the system is far away  from the critical point,
say 
\bee
g(t)>1+\hat g,
\eee
where $\hat g$ will be discussed below. 
It is assumed in this regime that the system's wave-function $|\psi(t)\rangle\simeq e^{i\phi(t)}|g(t)\rangle$, where 
 $|g(t)\rangle$ denotes the instantaneous ground state of $\hat H(g(t))$ (\ref{HIsing}) and $\phi(t)$ is the global phase.
Such dynamics is possible when the gap in the excitation spectrum is large enough.
\item The impulse regime happens around the critical point,  
\bee
\BBB{g(t)-1} < \hat g,
\eee
and the wave-function, apart from the overall phase factor, 
is assumed to be constant:
$|\psi(t)\rangle\simeq e^{i\phi(t)}|1+\hat g\rangle$. This happens when the gap in the excitation spectrum
becomes too small to support adiabatic dynamics {\it at the given quench time}.  
\item The second adiabatic regime happens when 
\bee
g(t)< 1-\hat g,
\eee
where it is assumed that no additional excitations are created by the quench,
i.e., $\BBB{\langle g(t)|\psi(t)\rangle}={\rm const}$.
\end{itemize}
These three stages of evolution are easily seen on Fig. \ref{fig_prob}a, where
the squared overlap of the wave function $|\psi(t)\rangle$ onto the instantaneous 
ground state $|g(t)\rangle$ of the Hamiltonian (\ref{HIsing}) is plotted
for a typical time evolution. We mention in passing that we 
have done the numerics for this section 
in the same way as in our recent Ref.  \cite{BDcd} discussing
the counterdiabatic dynamics of the Ising chain \cite{AdolfoPRL2012}.

The size of the impulse regime is governed by the gap in the excitation spectrum 
and the rate of driving. It is estimated by solving the following
equation (see e.g. Refs.  \cite{BDPRL2005}, \cite{DornerPRL2005}, and  \cite{BDPRL2007} for
different versions of this simple estimation and the physical reasons
supporting it)
\bee
\frac{1}{{\rm gap}(1\pm\hat g)}\sim\left.\frac{\rm gap}{\frac{d}{dt}\rm
gap}\right|_{g=1\pm\hat g}.
\eee

Assuming that the system is infinite, the relevant energy  gap in the quantum Ising model
is given here  by $2|g-1|$ for $g\ge0$ that we consider in this section. This 
gives us 
\be
\hat g \sim \frac{1}{\sqrt{\tau_Q}}.
\label{hatg}
\ee
It should be stressed that there is no sharp splitting between the adiabatic
and impulse dynamics. Instead, there is a crossover and Eq. (\ref{hatg}) provides 
the scaling of its location with the quench time $\tau_Q$.
Such estimation is meaningful for slow quenches only, $\tau_Q\gg1$, which can be numerically checked.

The quench time, however, cannot be too large in a finite Ising chain if we
want to see the impulse regime. Indeed, given the fact that the  gap at the 
critical point scales as $1/N$, one can always drive the system so slow that 
it will adiabatically cross the critical point (its excitation will be then 
exponentially
small in the quench time \cite{Marek_decoh}, which we illustrate in Fig. \ref{fig_prob}b). 
Note that the
gap in this section refers to the difference between the eigenenergy of 
the ground state and the lowest excited state that can be populated during the evolution.
It shall not be confused with the gap between the positive and negative parity
subspaces studied in Sec. \ref{sec_Qu}; the parity is conserved during the time
evolution. 

We expect that the finite-size effects are negligible during a non-equilibrium quench, when the 
spin-spin correlations in the system's wave-function $|\psi(t)\rangle$ decay on
a length scale much smaller than the system size. Within the adiabatic-impulse approximation
this length scale is upper bounded by the equilibrium correlation length (\ref{xii}) taken at 
$1+\hat g$. Thus, it is reasonable to expect that the finite-size effects are negligible when 
\be
N\gg\xi\B{1+\hat g}\sim\sqrt{\tau_Q},
\label{Ngg}
\ee
which we assume below. This educated guess  is rigorously discussed in
Refs.  \cite{JacekPRL} and  \cite{Marek_decoh}.

Using the 
adiabatic-impulse approximation, we estimate that the probability $p_{GS}$ of finding the 
system in the instantaneous ground state  away from the critical point, i.e.,
when $|g-1|\gg\hat g$, is  
\begin{align}
\label{ghj}
&p_{GS}=|\langle 1-\hat g|1+\hat g\rangle|^2 =  F^2(1,\delta),\\
&\delta=\hat g\sim \frac{1}{\sqrt{\tau_Q}}.
\label{mnb}
\end{align}
Combining Eqs. (\ref{Ngg}) and (\ref{mnb}), we see that the condition (\ref{poi}) is satisfied. 
Therefore, the thermodynamic expression (\ref{lnFc1}) for fidelity should be used in Eq. (\ref{ghj}), 
which leads us to 
\begin{equation}
p_{GS}=2\exp\B{-N\times \frac{\rm const}{\sqrt{\tau_Q}}},
\label{KZP}
\end{equation}
where the prefactor of $2$ comes from the subleading term in Eq.
(\ref{lnFc1}).

\begin{figure}[t]
\begin{center}
\includegraphics[width=4.5in]{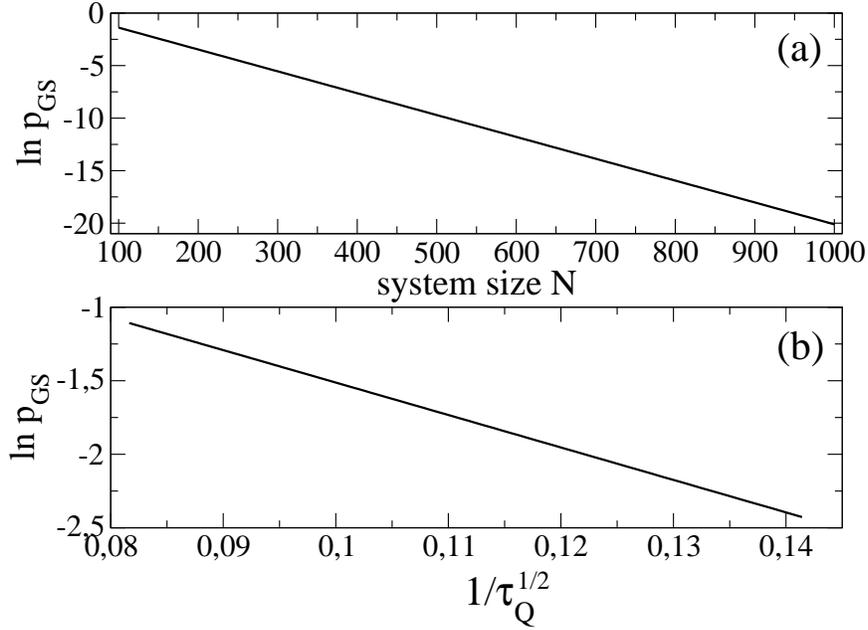}
\end{center}
\caption{Logarithm of the probability of finding the Ising chain 
in the ground state after the quench; see the discussion around  Eq.
(\ref{lnpgs}) for the details.
}
\label{fig_fity_dynamics}
\end{figure}

This simple prediction can be compared to numerics. To do so, we employ 
the quench protocol  (\ref{godt}). The evolution starts from the paramagnetic
phase  ground state at the magnetic field $g(t=-5\tau_Q)=5$ and it ends in the
ferromagnetic phase at $g(t=0)=0$. 
To simplify the comparison between the theory
and numerics, we take the logarithm of Eq. (\ref{KZP}) getting
\be
\begin{aligned}
&\ln\B{p_{GS}}= \ln\B{2} - N\times\frac{\rm const}{\sqrt{\tau_Q}},\\
&\ln\B{2}=0.693147\cdots.
\end{aligned}
\label{lnpgs}
\ee

First, we choose the quench time $\tau_Q=50$ and change the system size
$N$ in the range $[100,1000]$.  
The standard linear fit to the numerics from  Fig. \ref{fig_fity_dynamics}a yields:
\bee
\ln\B{p_{GS}}=0.693146(1) - 0.020800730(1)N.
\eee
The linear dependance of $\ln\B{p_{GS}}$ on the system size $N$ is nicely confirmed by the fit, and we note
that the fitted  intercept  surprisingly well agrees with the theoretical
one. 

Second, the system size is set to $N=150$ and the quench time $\tau_Q$ is varied in the range $[50,150]$.
The linear fit to the numerics from Fig. \ref{fig_fity_dynamics}b gives
\bee
\ln\B{p_{GS}}=0.6940(1)-\frac{22.065(1)}{\sqrt{\tau_Q}}.
\eee
The intercept again agrees surprisingly well with the theoretical prediction and
the linearity of $\ln\B{p_{GS}}$  in the inverse square root of the quench time is
nicely confirmed. 

The fitted lines are not depicted on Figs. \ref{fig_fity_dynamics}a
and \ref{fig_fity_dynamics}b because they are practically indistinguishable
from the numerics. It is worth to stress that for the system sizes considered 
in these numerical studies the inclusion of the subleading term to fidelity
per lattice site (\ref{lnFc1}) is crucial  for getting a good agreement
between the theoretical prediction for $p_{GS}$ and numerics.

Finally, we mention that a more general result can be obtained if we assume
that the gap closes at the critical point in the infinite system as $\BBB{g-g_c}^{z\nu}$,
where $z$ and $\nu$ are the universal critical exponents. Repeating the steps outlined
above one gets \cite{BDfid2}
\begin{equation}
p_{GS} \sim \exp\B{-N\times \frac{\rm const}{\tau_Q^{d\nu/(1+z\nu)}}},
\label{KZP1}
\end{equation}
where $d$ stands for the dimensionality of the system. The Ising result is
recovered for $z=\nu=d=1$.

The results (\ref{KZP}) and (\ref{KZP1}) have 
simple interpretation in the context of symmetry breaking phase transitions, which
in the quantum Ising model happen when the system is driven from the
paramagnetic to the ferromagnetic phase.
As predicted by the Kibble-Zurek theory \cite{Adolfo}, such transitions lead to 
the creation 
of topological defects whose density scales with the quench time 
as $1/\tau_Q^{d\nu/(1+z\nu)}$.
Thus, the probability of finding the system after the quench in its
instantaneous ground state is exponential 
in the number of topological  defects created during the non-adiabatic
crossing of the quantum critical point.

\section{Fidelity in central spin systems}
\label{sec_Cen}
It turns out that knowledge of fidelity helps in the studies of the central
spin-1/2 coupled to the Ising chain \cite{BD_dec,BDSciRep2012}. 
The Hamiltonian of such a  system reads 
\bee
\hat H(g)-\delta\sum_{j=1}^N\sigma _{j}^{z}\sigma^{z}_{\cal S},
\eee
where $\hat H(g)$ is given by Eq. (\ref{HIsing}), $\sigma^z_{\cal S}$ is the Pauli matrix of the central spin,
and  $\delta$ is the central spin--environment coupling. Such a model is known as the central spin model because
of the uniform coupling of the central spin to its ``environment''. 
Without going into details, we will mention two problems, where  the 
expressions for fidelity from 
Secs. \ref{sec_Sus} and \ref{sec_The} can be used.

{\it Critical dynamics of decoherence.} Suppose that the magnetic field is quenched by the protocol 
(\ref{godt}). The system is initially in the product state
\be
(c_{+} \left |\uparrow \right \rangle +c_{-}\left |\downarrow \right\rangle)\otimes
\left|g(t=-\infty)\right\rangle,
\label{productstate}
\ee
where the central spin is in an arbitrary superposition of its up/down states and the Ising chain 
is in the ground state $\left|g\right\rangle$ of $\hat H(g)$.
The idea now is to study the decoherence of the central spin due to the presence of
the many-body Ising environment undergoing a non-equilibrium quench across the
critical point(s). 
As is explained in Ref.  \cite{BD_dec}, the knowledge of
fidelity $F(g,\delta)$ is crucial for the explanation of the decoherence rate
of the central spin in this non-equilibrium system.

{\it Magnetic Schr\"odinger's cats.}
We consider now the same system as the one described above and assume that
the Ising chain is adiabatically driven towards the critical point, 
where the time variation of the magnetic field
(\ref{godt}) stops.
Assuming that the evolution starts from the initial state (\ref{productstate})
and that $\tau_Q\gg N^2$,  the final state of the system will be 
approximately  given by 
\bee
c_-e^{i\phi_-}\left| \downarrow \right\rangle \otimes |g-\delta\rangle+
c_{+}e^{i\phi_+} \left| \uparrow \right\rangle \otimes |g+\delta\rangle,
\eee
where $\phi_\pm$ stand for the phases picked by the system during the
evolution \cite{BDSciRep2012}.
Assuming that $g-\delta < g_c < g+\delta$,  
$|g-\delta\rangle$ and $|g+\delta\rangle$  are the ferromagnetic and paramagnetic ground states of the Ising
Hamiltonian. 
This state is the magnetic Schr\"odinger's cat
state. The dead/alive cat states are the ferromagnetic/paramagnetic
ground states. The  state of the radioactive atom that has/has not decayed 
is played by the up/down states of the central spin. It is not, however,
the ordinary Schr\"odinger's cat because the two cat states are not orthogonal,
\bee
\BBB{\langle g-\delta|g+\delta\rangle} = F(g,\delta) \neq 0.
\eee
It would be now certainly interesting to understand 
  how the rate of decoherence of such a superposition state depends on the similarity 
between the two cat states, i.e., their  fidelity. 
It is perhaps worth to
stress that the experimental studies of the decoherence of ``simpler''
superpositions have been recognized \cite{DJWineland,Haroche} by the Nobel  committee in 2012, so it
would be certainly very exciting to experimentally quantify the decoherence rate of the
 magnetic  Schr\"odinger's cats! For other applications of fidelity in this
 context see Ref.  \cite{BDSciRep2012}. For similar ideas in the cold ion
 and cold atom systems  see Refs.  \cite{Morigi} and  \cite{Ritsch},
 respectively.

\section{Summary}
\label{sec_Su}
We finish these lecture notes by quoting P. W. Anderson \cite{Anderson1967}
``{\it While wave functions and overlap integrals
are often of little consequence in many-body
systems, this one is at least related to the response
to a sudden application of the potential, and indicates  that response involves
only the emission of low-energy excitations
into the continuum, as well as that the truly
adiabatic application of such a potential to such
a system is impossible}''. We hope that these lecture notes provide further evidence that the 
wave functions and overlap integrals are indeed very useful in the studies of many-body quantum 
systems.

\section*{Acknowledgments}
This work has been supported by the Polish National Science Centre (NCN) grant DEC-2013/09/B/ST3/00239.
I would like to thank Marek Rams for  the collaboration on the topics 
discussed in these lecture notes and for insightful comments about this manuscript.


\end{document}